\documentclass[intlimits,twoside,a4paper]{article}

\usepackage{amsmath,amssymb}
\usepackage{graphicx,psfrag}

\usepackage[T2A]{fontenc}
\usepackage[cp1251]{inputenc}

\usepackage[eqsecnum]{cmpj3}

\RequirePackage{color}

\issue{2020}{23}{2}{23001}
\doinumber{10.5488/CMP.23.23001}

\title[Crossing borders in the 19th century and now --- two examples of weaving a scientific network]%
{Crossing borders in the 19th century and now --- two examples of weaving a scientific network}
\author[R. Folk, Yu. Holovatch]{R. Folk\refaddr{label1}, Yu. Holovatch\refaddr{label2,label3,label4} }
\addresses{
\addr{label1}Institute for Theoretical Physics, Johannes Kepler University Linz, 4040 Linz, Austria 
\addr{label2}Institute for Condensed Matter Physics of the National Academy of Sciences of Ukraine, 79011 Lviv, Ukraine 
\addr{label3}$\mathbb{L}^4$ Collaboration \& Doctoral College for the Statistical Physics of Complex Systems, 
Leipzig-Lorraine-Lviv-Coventry, Europe 
\addr{label4}Centre for Fluid and Complex Systems, Coventry University, Coventry, 
CV1 5FB, United Kingdom}

\date{Received January 19, 2020, in final form February 17, 2020}

\begin{document}
\maketitle
\begin{abstract}
Scientific research is and was at all times a transnational (global) activity. In this respect, it crosses several borders: national, cultural, 
and ideological. Even in times when physical borders separated the scientific community, scientists kept  their minds open to the ideas created beyond the walls and tried to communicate despite all the obstacles. An example of such activities in the field of physics is the travel in the year 
1838 of a group of three 
scientists through the Western Europe: Andreas Ettingshausen (professor at the University of Vienna), August Kunzek (professor at the University of Lviv) 
and P. Marian Koller (director of the observatory in Chremsminster, Upper Austria).  
 $155$ years later a vivid scientific exchange  began between physicists from Austria and Ukraine,  in particular, between the Institute for Condensed Matter Physics of the
National Academy of Sciences of Ukraine in Lviv and the Institute for Theoretical Physics of Johannes Kepler University Linz. 
This became possible due to the programs financed by national institutions, but it had its scientific background in already knotted historic 
scientific networks, when Lviv was an international center of mathematics and in Vienna the `School of Statistical Thought' arose. 
Due to the new collaboration, after the breakup of the Soviet Union, Ukraine became the first country to join the Middle European Cooperation 
in Statistical Physics (MECO) founded in the early 1970s with the aim of bridging the gap between scientists from the Eastern and Western 
parts of Europe separated by the iron curtain. 
In this paper, we discuss the above examples of scientific cooperation pursuing several goals: to record the less known facts from
the history of science in a general culturological context, to trace the rise of studies
that in due time resulted in an emergence of statistical and condensed matter physics as well as
to follow the development of multilayer networking structures that join scientists and enable their 
research. It is our pleasure to submit this paper to the Festschrift devoted to the 60th birthday
of a renowned physicist, our good colleague and friend Ihor Mryglod. In fact, his activities 
contributed a lot into strengthening the networks we describe in this paper.


\end{abstract}

\section{Introduction}\label{I}

\hspace{25em} 
{\em Science is rooted in conversation.}

\hspace{25em} 
Werner Heisenberg \cite{heisenberg}.
\\

\hspace{25em} 
{\em Knowledge propagates.}

\hspace{25em} 
Stuart Kauffman \cite{kauffman}.
\\

Apart from exceptional, nevertheless famous examples in the history of scientists trying to hide their research,
nowadays the community of scientists would agree with Heisenberg's opinion cited above~\cite{liberwolf}. However, private communication is 
only one (maybe the most intimate) of the ways to spread the ideas and knowledge. There are other networks permitting propagation of knowledge, the most 
relevant being the educational networks at different levels. An important one here is the academic geneological network, then come the networking due to publications, 
talks on conferences, free or forced migrations of scientists during their carrier, connections due to national and international programs, organizing the 
transfer between disciplines and between research and application. Each of these processes  is nowadays the object of interdisciplinary research
field of the {\em science of science}, see for example  
\cite{stanley} and \cite{shietal}. Moreover, as far as the creation of networks cost money their supporters are interested in the evaluation of these 
networks \cite{resnetw,Berche16}.

Here, we will consider two examples separated by a time period of 155 years. Both of them are taken 
from the field of physics and concern the communication between scientists from the regions of the origin
of the authors of this paper. Doing so, we pursue several goals: to record the less known facts from
the history of science in a general culturological context, to trace the rise of studies
that in due time resulted in an emergence of statistical and condensed matter physics,
to follow the development of multilayer networking structures that join scientists and enable their 
research. It is our pleasure to submit this paper to the Festschrift devoted to the 60th birthday
of a renowned physicist, our good colleague and friend Ihor Mryglod. In fact, his activities 
contributed a lot into strengthening the networks we will speak about in this paper.

The rest of the paper is organized as follows. In the next section \ref{II} we describe more in
detail the above two examples of scientific cooperation: a travel of three scientists (from Vienna,
Lviv and Chremsminster near Linz) in 1838 through the leading centers of European scientific thought and a collaboration 
between the {Institute of Theoretical Physics of the Johannes Kepler University in Linz and the Institute for Condensed Matter Physics in Lviv} 
that started 155 years later. Section \ref{III} collects some facts about the developments in
statistical physics related to the above collaboration and section \ref{IV} contains some {general reflections}.

\section{Scientific travelling {and research cooperation}}\label{II}

\begin{figure}[!b]
\centering\includegraphics[height=5.8cm]{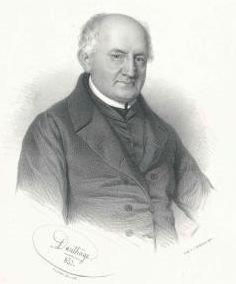}\includegraphics[height=5.5cm]{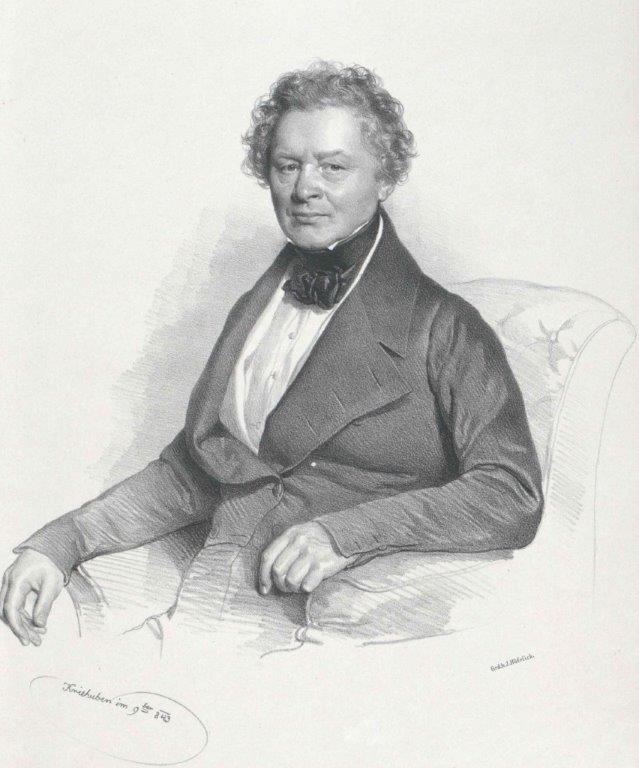}\includegraphics[height=5.5cm]{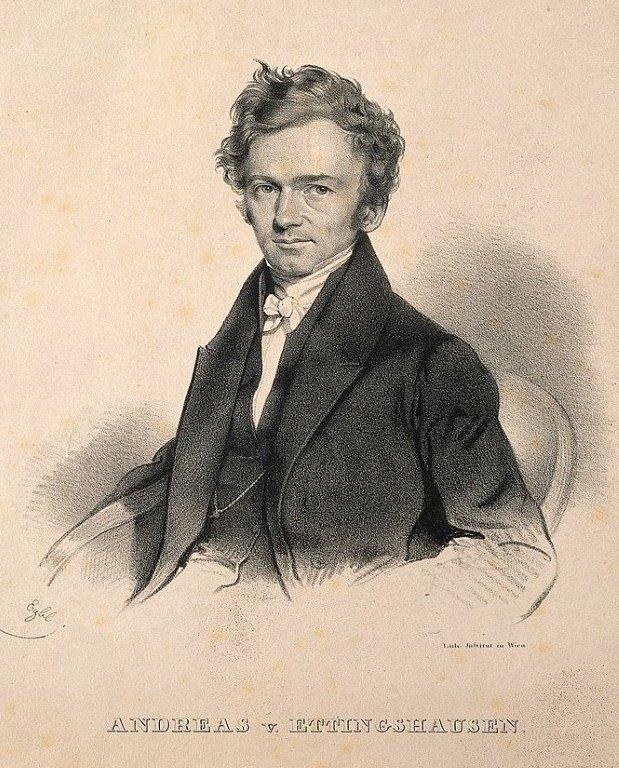}
\caption{The three travelling companions: P. Marian Koller, Prof. August Kunzek and Prof. Andreas Ettingshausen. 
 \copyright  Bildarchiv Austria \"ONB.
\label{fig1}}
\end{figure}

The authors of this paper are from two different institutions: {the Institute of Theoretical Physics (ITP) of the Faculty of Engineering \&
Natural Sciences of the Johannes Kepler University Linz 
in Austria} and {the} Institute for Condensed Matter Physics of the National Academy of Sciences of Ukraine (ICMP) in Lviv.
It so happened that in 2019 both institutions celebrate their jubilees:  both ITP as one of the Faculty institutions and the first department of ICMP
were founded in 1969. In the meantime, a tight collaboration in the field of statistical and condensed matter physics
has been established between our institutions. This was initiated by our first common projects back in 1993 and 
became possible due to the programs financed by national institutions. Obviously, such a collaboration
has its scientific background in the already knotted historic 
scientific networks, when Lviv was an international center of mathematics~\cite{Math,Lviv_math1,Lviv_math2}  and when the `School of Statistical Thought' arose in Vienna \cite{montroll}.

\begin{figure}[!t]
\centering
\includegraphics[width=10.8cm]{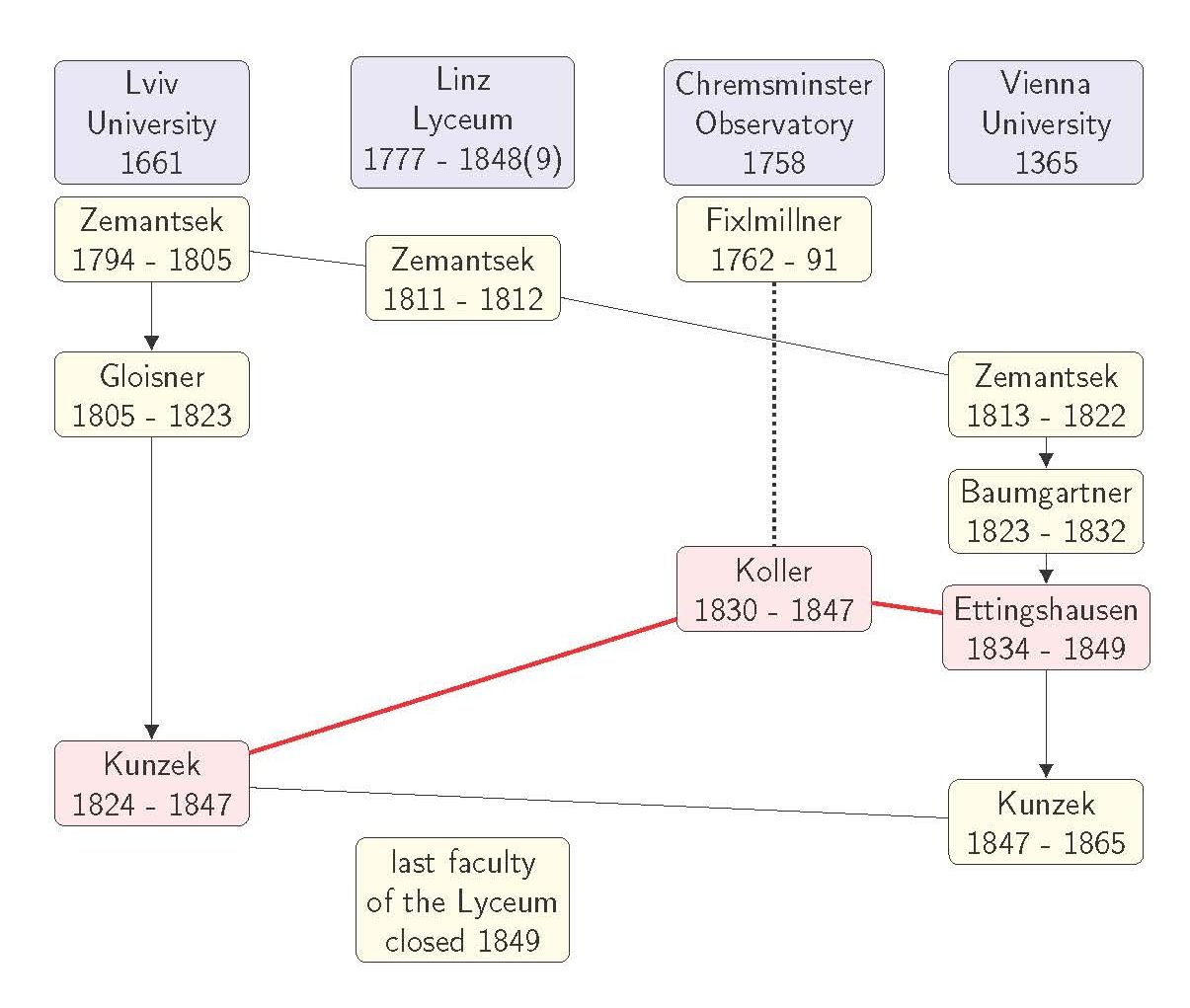}
\caption{(Colour online) Positions at different institutions: arrows denote followers in the position (chair at universities); 
lines: temporary migration, dashed lines: not all followers are named; red line: participants in the scientific journey 1838 (see figure~\ref{fig1} and 
the text). \label{fig2}}
  \end{figure}
  
To shed more light on the origin, development and possible prospects of common scientific inquiries,
we decided as a case study to consider more scrupulously two examples of cooperation between scientists from
our regions. The first example is the travel in the year 1838 of a group of three 
scientists through Western Europe. These were  Andreas Ettingshausen (professor at the University of Vienna), August Kunzek 
(professor at the University in Lviv) 
and Pater Marian Koller (director of the observatory in Chremsminster, near Linz in Upper Austria). The second example
is given by the above mentioned cooperation between {the ITP and the ICMP} that began 155 years later.

As it becomes apparent from the further account, heroes of our stories had connections to different institutions,
that emerged and disappeared in the course of their life (see figure~\ref{fig2}). Moreover, the countries disappeared and {reappeared} too, eliminating
old borders and establishing new ones. The city of Lviv (Lw\'ow in Polish and Lemberg in German) had 
its university since 1661, presently the Ivan Franko National University of Lviv. 
Linz  attained its university in 1966 first as a University of Social and Economic Sciences, later being enlarged by further faculties.
Earlier, in Linz at the protestant ``Landschaftsschule''
 Johannes Kepler taught mathematics, and afterwards in 1777 the catholic Lyceum was founded here 
(both schools are considered as forerunners of the university). Moreover, the nearby  monastery  of Chremsminster was an important 
educational center with its schools and a research center of astronomy and natural sciences. Its astronomical and geophysical observatory 
(``Mathematical Tower'') was 
built in 1749--1756  and was well connected to other observatories in Europe. 

There is no direct connection between the two examples of cooperation of the scientists that we discuss in this section.
The common feature, however, is the openness to new ideas and the basis of a common network apart from the specific topic. 
This could be seen, regarding the first example, as an academic network in the  Habsburg university landscape to which the scientists 
belonged and, regarding the second case, the European academic background which developed during the period of cooperation. The field of 
critical phenomena, properties of condensed systems and the theory of statistical mechanics were developed during the time 
frame in between the two examples.

%\newpage

\subsection{Three men cross Europe {in the year 1838}}\label{II.1}

The plan for this {journey \cite{amand}} came from Andreas v. Ettingshausen\footnote{He was pupil of the gymnasium at the monastery Chremsminster.}  
(1796--1878) professor for physics at the University of Vienna and should include Pater Marian Koller  (1792--1866) 
director\footnote{He also taught  physics at the monasterial gymnasium and was director of the Physical Cabinet at the observatory \cite{felloecker}.} 
of the observatory of the monastery Chremsminster and Andreas Baumgartner {(1793--1865)} also professor of the University of Vienna. 
The main purpose was to collect maximum information on the current scientific projects, new instruments and teaching facilities. 
The ranges of interests were quite broad according to the different inclinations of the participants.
The preparation started almost a year earlier in order to get an official permission and financing for the journey.  
Due to the health problems, Andreas Baumgartner was forced already in 1833 to reduce his teaching activities,
therefore he had to withdraw his participation in the journey and  was replaced 
by August Kunzek (1795--1865), professor of physics at the University of Lviv.\footnote{Both Baumgartner and Kunzek taught 
at the Lyceum in Olomouc before they became professors at the respective universities.} In figure~\ref{fig2} we show
a diagram sketching the succession and mobility of scientists in the institutions under discussion and displaying
in this way the continuity of academic tradition and knowledge propagation.

The journey of three colleagues-scientists through Europe took place in 1838. P. Koller regularly reported pieces of news from different places 
to his assistant in the observatory, P. Reslhuber  \cite{felloecker}. A detailed description of the journey can be found in P. Reslhuber's 
biography of P. Koller \cite{reslhuber}.  The stay of these scientists in Berlin is also 
documented in the recollections of Rudolf Wolf \cite{wolf}. We collect some data about the journey in table~\ref{tab1},
presenting the time-table, names of the cities visited as well as of the scientists and instrument makers they met there.
When the journey was over, the Austrian newspaper {\it Der Adler} of November 6, 1838 wrote (p. 1045): ``\ldots A few days ago, 
the professor of physics in Lemberg, Dr. August Kunzek, returned from his scientific journey through Germany, Belgium, England and 
France, which he undertook with the professors of physics A. von Ettingshausen and Marian Koller. The professors of the 
local philosophical school {arranged} a feast in the  {casino pub} to receive him, in order to testify their participation 
in the great scientific exploitation of the learned travellers, and the promotion of science thereby 
effected\ldots''.\footnote{Translated from German by R.F.}

Kunzek travelled abroad at his own expense, especially to Germany, France, and England, where he expanded his  knowledge by visiting 
scientific institutes, museums, and laboratories. Many  achievements in Lviv had their basis on these visits. When the technical academy 
was to be established in Lviv, he was also entrusted by the Government with the drafting of a plan to organize it, and in the following 
his proposals were 
indeed realized.\footnote{The k. \& k. (kaiserlich und k\"oniglich, i.e., Imperial and Royal) Real School was opened in Lviv in 1816. In 1825, according to the 
Royal Decree of the Austrian Emperor Franz I, the three-level k. \& k.
  Real School was reorganized into the  k. \& k. School of Technical Sciences and Trade.
In 1835 it turned into the k. \& k. Real-Trade Academy, it was one of the first academic technical schools in Europe and the first in Ukraine. 
The academy was renamed as Polytechnic School and included in the academic schools of the Austro-Hungarian Empire.}
When in 1844, the  k. \& k. Technical Academy with technical and trade departments was opened in Lviv,
August Kunzek was beneath 14 persons, who applied for the position of its director, but the position went to 
Florian Schindler, teacher at the Joanneum in Graz (\cite{ditchen1} p. 103). In 1847 Kunzek obtained the chair of physics and applied mathematics 
at the University of Vienna \cite{surman}. Boltzmann visited the following lectures by Kunzek: WS 1863/64 Light and heat; SS 1864 Statics of liquid bodies; 
WS 1864/65 Magnetism \cite{katgraz}.

\begin{table}[!t]
\caption{(Colour online) Map and time-table of the scientific journey and the scientist and instrument makers met at  
different stations. In blue, the travel to Vienna, in black and red, different stages of the journey. \label{tab1}}
\vspace{2ex}
\centering\includegraphics[width=14.5cm]{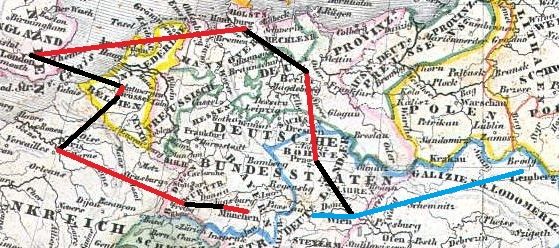}\vspace{0.2cm}
{
\begin{tabular}{|c|c|c|} \hline\hline
Date & Place & Persons  \\ \hline\hline
18.07  &Prague&F. Hessler, K. Wersin, Ch. Doppler \\ \hline
&&J.G. Galle, J.H.A. Oertling, F.W. Schiek, \\
25.07 &Berlin&W. Hirschmann, J.H. M\"adler,  \\
&&E. Mitscherlich, R. Rie\ss, G. Magnus\\ \hline
27.07 &Hamburg &K. R\"umker, H.Ch. Schuhmacher, J. Herschel, \\
&& H. Kessels, Repsold, A.C. Petersen  \\ \hline
 & & J. Herschel, F. Baily, Ch. Babage, \\  
03.08&London& J.D. Roberton, G. Dollond, M. Faraday, \\
&&S.W. Stratford, G. Airy\\ \hline
 &Antwerpen &visit to town only \\ \hline
21.08 &Br\"ussel &J. Quetelet \\ \hline
&&H.-P. Gambey, L.C. Breguet, F. Arago,   \\
 & &A. Bouvard, F. Savary, S.D. Poisson,  \\
24.08 &Paris &Ch.-F. Sturm, E. Chevreul, A. Dumeril, \\
 & &A. Brogniart, J.B.B. St. Vincent, J. Babinet, \\
 & &{\bf Ch.C. de la Tour}, A. Humboldt, J. P\'eclet, \\
 & &C. Pouillet, A. Cauchy, Ch. Chevalier \\ \hline
 && L. von Buch, W. Buckland, K.F. Martius, \\
16.09 &Freiburg &Ch.F. Sch\"onbein, G.W. Munke,  \\
&& G. Osann, W. Eisenlohr\\ \hline
21.09 &Augsburg& Stark \\ \hline
27.09 &M\"unchen & Steinheil, Lamont \\ \hline\hline
\end{tabular}}
\end{table}

One of the priorities of the journey was to visit the observatories to get information in the field of astronomy, but also other fields of research such as 
meteorology, and primarily geomagnetism, were of interest. Meteorological and geomagnetic measurements were made since 
the 18th century in the observatory and in the year 1839,  due to the cooperation with Karl Kreil (1798--1862) (in Prague), 
the measurement station of the observatory\footnote{A Gauss' magnetometer  was purchased.} became a part of the international 
network of the `G\"ottinger Magnetischen Vereins' \cite{zamg}. 
In addition, in his publication {\it Ueber den Gang der W\"arme in \"Osterreich ob der Enns}~\cite{koller} P.~Marian Koller 
``\ldots imparted the farsighted lesson that the future of climatological research would require cooperation on a large scale, 
following the example set by Humboldt (whom he met in Paris, see table~\ref{tab1}) and Gauss for geomagnetism\ldots'' (see \cite{coen} chapter 3, p. 68).

Let us point out {another French scientist they met in Paris}, connected to the research, which was 155~years later a common topic of 
cooperation between University of Linz and the ICMP in Lviv. {This was at  the meeting of the Academy} where they 
got to know Charles Cagniard de la Tour (1777--1859), who in 1822 discovered special effects in the liquids at a certain point 
(in the temperature--pressure plane), later named as a critical point  \cite{BHK1,BHK2}. The behaviour of matter near 
such a peculiar point was then named a critical 
phenomenon and opened up a large field of the research in physics spreading out to other fields even outside natural sciences such  as 
economics, sociology or even humanities \cite{Holovatch17,Ising17}. However, at the time of the journey of the three 
scientists, other topics were in discussion from such fields as astronomy, meteorology, and most notably optical, electrical 
and magnetic phenomena.  In 1865 they led to Maxwell's electrodynamics unifying those three phenomena in one field theory. 

\begin{figure}[!t]
\centering\includegraphics[width=12.5cm]{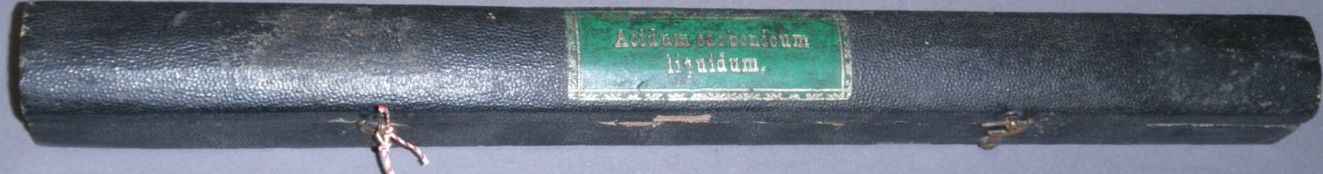}\\\vspace{0.2cm}\includegraphics[width=12.5cm]{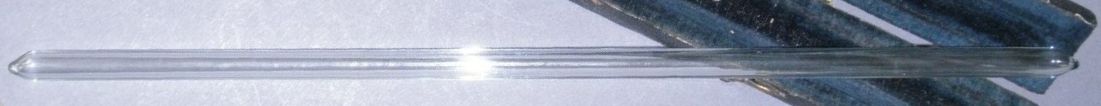}
\caption{\label{natterer} (Colour online) Natterer tube with CO$_2$ used in lectures to show critical opalescence. $\quad$ \copyright~Collection in 
the Observatory of the Monastery Chremsminster.}
\end{figure}

In 1854 in Vienna Johannes Natterer tried to liquefy the air by increasing the pressure but failed. The reason was that he 
was unaware of the concept of the critical point, an explanation was first given by Dmitri Mendeleev  theory of the absolute 
boiling point of liquids and Thomas Andrews experiments with CO$_2$ gas \cite{almqvist,brenni}.  In order to liquefy, the 
temperature of the gas should be below the critical temperature. It happened only by chance to succeed by increasing 
the pressure, while for the cases where the method failed, the gases were named permanent gases. 
In 1858 a device of Natterer  for demonstrating critical phenomena arrived at Chremsminster. 
The famous Natterer tube (see figure~\ref{natterer}) was mentioned in the publications of Andrews in 1869 and Smoluchowski in 1911. It shows the 
disappearance of the meniscus between the gaseous and liquid phase at the critical temperature by increasing the temperature 
from lower temperatures and the critical opalescence by reducing the temperature coming from higher temperatures to the 
critical one.

\subsection{Science as a collective enterprise}\label{II.2}

Organization of scientific research changed considerably in between the period of Koller's, Kunzek's and Ettingshausen's travel and the 
beginning of the Linz-Lviv cooperation in 1993. Derek J. de Solla Price characterized it shortly \cite{price} as a change from {\it Little Science} 
to {\it Big Science} or, as one may say, {\it from the study room to large-scale research}. The main changes began about 1900 but especially 
after the two World Wars: (1) an increase in the number of researchers and publications\footnote{Apart from a short reduction due to the 
boycott against central scientists \cite{beaver,iaria}.}, (2) an increase of publications with coauthors, (3) a change in the local organization 
of research from individuals to teams,  (4) globally, an increased cooperation in common research programs between different countries and (5)~a spread of knowledge due to displacement of scientists because of political reasons but above all because of antisemitism.\footnote{It is the task 
of the {\it science of science} to quantify this development and create models to give predictions regarding the success of these processes 
\cite{fortunato}.} Together with this, the possibilities of individual communication between scientists  were speeded up 
from sending letters to sending  emails in the mid-1980-ies. 

On the one hand, the development was supported in Europe by  the formation of the European community, on the other hand, it was hindered by the Cold War 
and its manifestation, the {\it Iron Curtain}. Nevertheless, scientists tried to overcome these obstacles and to make the most of keeping 
in touch over the borders. Among numerous initiatives and forms of support for scientific collaboration, we would like to emphasize
several  initiatives that helped a lot in establishing and strengthening Linz-Lviv collaboration. These are the Middle European Cooperation 
in Statistical Physics (MECO) and European Cooperation in Science and Technology (COST).

\begin{figure}[!t]
\centering\includegraphics[width=10cm]{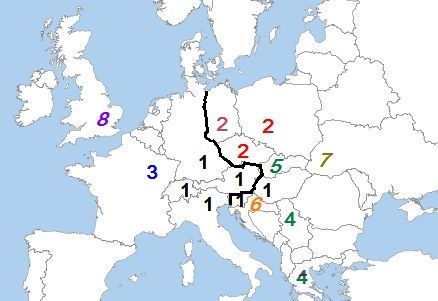}
\caption{\label{fig4} (Colour online) The map shows the countries that hosted MECO meetings in 1975--2019. The numbers show the order of the country 
coming to the advisory board. 
{The black line indicates the border between Western and Eastern countries; apart from the border between the former states of 
Yugoslavia, it formed the {\it iron curtain.}}}
\end{figure}

An idea to organize regular meetings of scientists from both sides of the Iron Curtain, where also young scientists could take part, 
emerged in the early 1970s \cite{MECO}. Very soon MECO became one of the most influential meeting places where 
the state-of-the-art ideas in statistical
and condensed matter physics were born and discussed. The topics of 
ferroelectricity and structural phase transitions, the 
soft mode  concept and the problem of a central peak, 
%new theoretical aspects of Wilson's and field-theoretical renormalization group 
new theoretical aspects of renormalization group
and much more were subjects of vivid and fruitful discussions at the first MECO meetings. These were exactly the topics that comprised
a core of Linz-Lviv collaboration. With a span of time, due to more and more active participation of Ukrainian scientists, Lviv
became the first city in the former Soviet Union that hosted  the MECO meeting too, see the sketch of widening of the MECO network in figure~\ref{fig4}.

The main goal of COST is to enable researchers from different fields and different countries to work together in open networks that transcend borders.
In particular, COST is funding Actions ---  a network dedicated to scientific collaboration, complementing national research funds. 

The first action in Physics was COST P1 --- Soft condensed matter which started in 1997 and ran until 2001. The set of subsequent Actions
centered upon applications of physical ideas in the fields far  beyond physics in its traditional sense, essentially contributed
to maintaining and strengthening Linz-Lviv cooperation too. In particular, these were: Physics of Risk (P10, 2005--2007), Physics of 
Competition and Conflicts (MP0801, 2008--2013), Analyzing the dynamics of information and knowledge landscapes (TD1210, 2013--2017).

\begin{figure}[!t]
\centering\includegraphics[height=4.7cm]{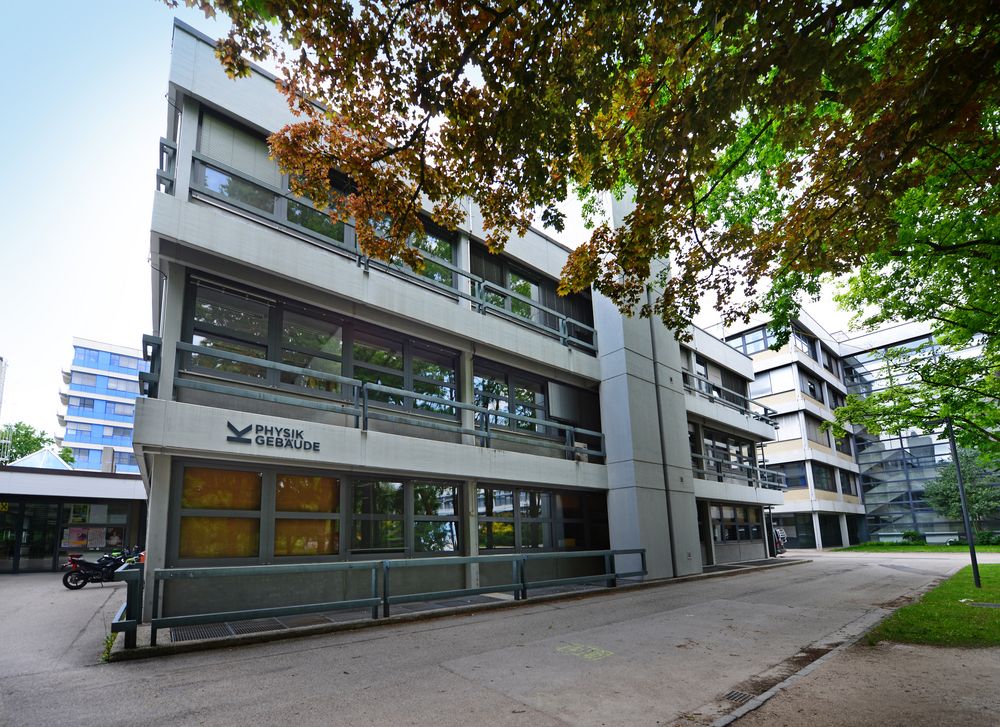}\hspace{0.1cm}\includegraphics[height=4.7cm]{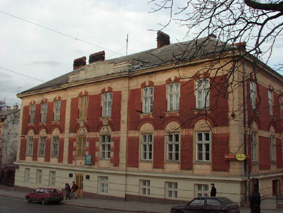}
\caption{\label{fig6} (Colour online) (a) The Physics building of the Johannes Kepler University, 69 Altenberger St., Linz, {containing the ITP}  was erected 
with the participation of the architect Artur Perotti 
(1920--1992). (b) The main ICMP building, located at the address 1 Svientsitskii St., Lviv (the former Instytutsjka St.), was built in 
1900 according to the project of  Ludwik Wierzbicki (1834--1912). The ICMP moved into this building in 1991.}
\end{figure}

The first contacts between ITP and ICMP (see figure~\ref{fig6} where we show current locations of these institutions)
started by personal visits in 1992--93 and were further developing during {the}
Ukrainian-French Symposium `Condensed Matter: Science \& Industry' (February 20--27, 1993, Lviv), Ukrainian-Polish and  
East-European Workshop on Ferroelectricity and Phase Transitions (September 18--24, 1994, Uzhhorod-V.~Remety) and 
many more meetings that were commonly attended or organized. The first common paper was written in 1995 \cite{paper} and since
then we have a good hundred of common publications both reporting original research, reviewing it and making it
popular to a wider community. Of course, a  measure for cooperation of two institutions, which is visible to 
all the community, is {the} number of papers with coauthorship from both institutions. However, there is, in addition, a less 
visible cooperation just from the exchange of information privately, in discussions sometimes visible in papers 
by acknowledgements. Moreover, authors of the common papers are themselves embedded in coauthorship networks leading 
to a larger network which is the basis for possible flow of information. All these we encountered in our collaboration.

All cooperations were financed by grants from of the Austrian `Wissenschaftsfonds' 
(FWF)\footnote{P19583-N20 {\it Critical phenomena in pure and disordered systems} (2007--2011), 
P18592-N20 {\it Phase transitions and correlations in complex fluids} (2006--2008),  
P16574-N08 {\it Critical phenomena in disordered systems} (2003--2007), P15247-N03 {\it Dynamics of complex fluids} (2001--2004), 12422-PHY (1997--2000).}, 
one project by The Anniversary Fund of the Oesterreichische Nationalbank (OeNB)\footnote{No.7694 {\it Critical phenomena} (1999--2003).}, 
one by the Ministry of Science, Research and Art\footnote{Grant {\it Microscopic theory of dynamic properties of magnetic liquids} (1994--1996).} 
shorter stays were supported by the OeAD. Mutual lecturing and guest professorships in Lviv and in Linz enabled a further development
of our contacts and involved young colleagues to future collaboration. Besides lecturing, we initiated translation of textbooks
written in a native language in one country in order to be used for students of other countries \cite{Iro_book}.  In table~\ref{tab2} we list the 
names of the colleagues
involved in this collaboration.\footnote{The poster below the table was produced by Olesya Mryglod, ICMP Lviv, to commemorate a benchmark
in the collaboration.} Although we do not specify participation of each of them in the specific part of collaboration projects, we think it 
is proper to note that it was Ihor Mryglod who made the first visit from the ICMP to the ITP,  followed by Yulian Vysochanskii  
from Uzhhorod University and by  numerous subsequent contacts.

\begin{table}[h]
\caption{\label{tab2} (Colour online) Group members involved in the collaboration from the ITP Linz and the ICMP Lviv apart from two exceptions: 
$^*$from the University of Uzhhorod,
$^{**}$from the Ivan Franko  National University of Lviv. The figure shows some of the travelling companions from Lviv that visited 
Johannes Kepler University Linz in the course of this collaboration.}
\vspace{2ex}
\centering {
\begin{tabular}{|c|c||c|c|}\hline
ITP Linz &ICMP Lviv&ITP Linz &ICMP Lviv\\ \hline
R. Folk & Yu.M. Vysochanskii$^*$&T.-C. Dinh& M. Dudka\\ \hline
G. Moser & I.M. Mryglod &I. Nasser& V. Blavats'ka \\ \hline
F. Schinagl  & Yu. Holovatch &A. Abdel-Hady& O. Prytula \\ \hline
H. Iro   & T. Yavors'kii$^{**}$ &G. Flossmann& V. Palchykov\\ \hline
W. Fenz &  I. Omelyan && \\ \hline
\end{tabular}}\vspace{0.2cm}
\includegraphics[width=13.0cm]{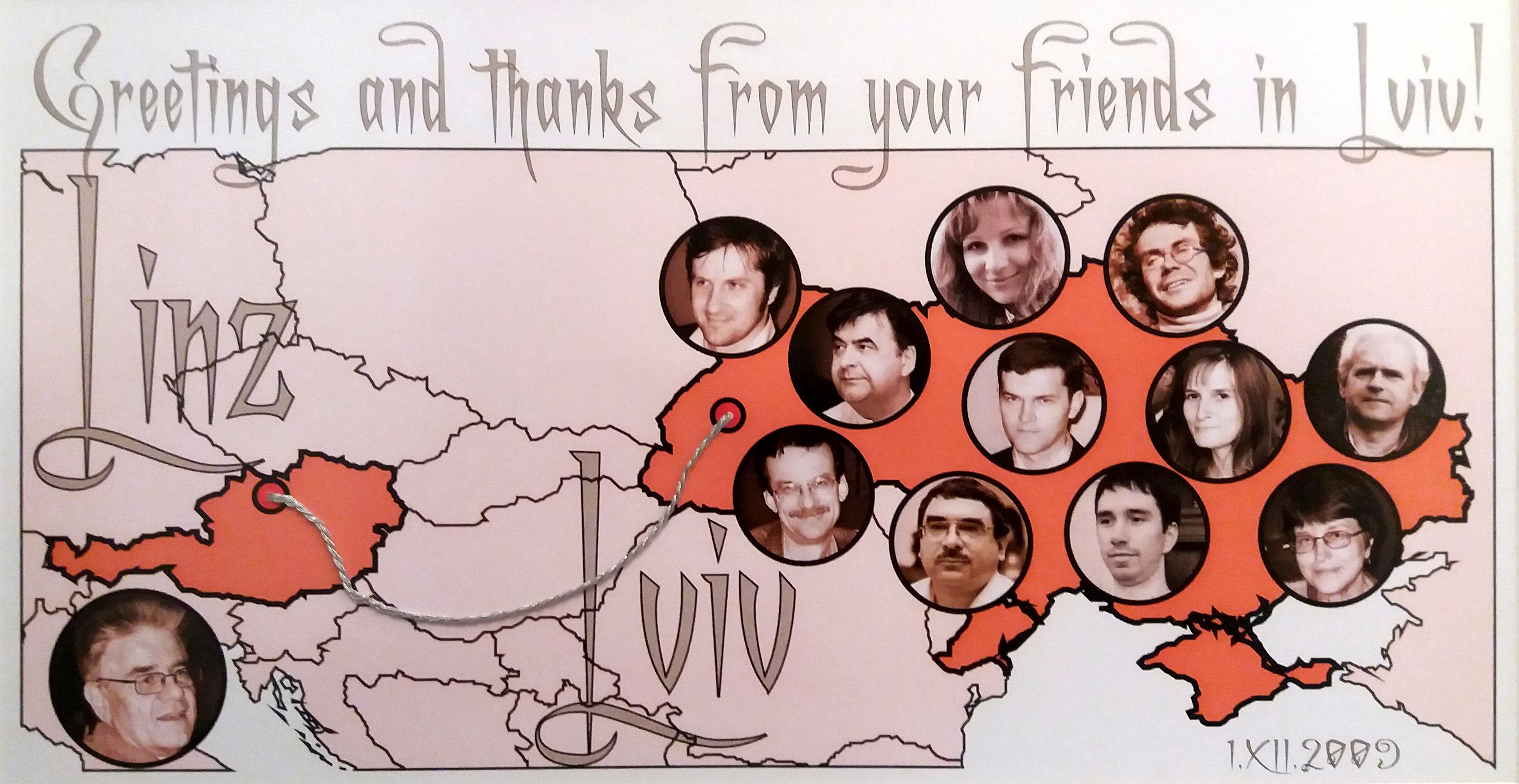}
\end{table}

The topics of our common research covered various fields of statistical and condensed matter physics. Beneath the objects of interest there 
were ferroelectrics, 
regular and structurally  disordered magnets, magnetic liquids, superconductors. We were interested in static and dynamic critical phenomena,   
crossover critical behaviour, self-organization and emergence of new features in complex systems.
And indeed, some of the phenomena that were in the core of our interest were the subject of discussion between travelling companions we spoke about in 
the former subsection. Moreover, the period in between gave rise to the new field of physics --- statistical physics, that enabled a
thorough theoretical analysis of these and other phenomena that occur in systems of many interacting particles. 
In our research, we  developed and learned the methods of renormalization
group, non-equilibrium statistical operator, functional integration and diagrammatic expansions, resummation of asymptotic series and creating 
algorithms for computer simulations. And again, some of the mathematics involved had to do with our predecessors from the regions
we discuss in this paper. In the next section we will give
some references to the key persons who created these fields and worked and lived in our regions.

\section{What happened in between?} \label{III}

In this section, we do not give a thorough and comprehensive description
of the rise and development of statistical physics ideas. We rather concentrate on
some personalities mentioning their affiliations, mobility and/or institutional
and intellectual connections. Doing so, we will emphasize here those facts that are 
mostly linked to {the ITP-ICMP} collaboration.

\subsection{Vienna school of statistical thought}

The first severe regress {in intensification of scientific connections} was the breakdown of the Habsburg 
Empire after the World War I. The developing academic network was destroyed and the survival of several universities was questioned. 
It is interesting that together with the demand to build Ukrainian university in Lviv in 1917, a similar application was placed for 
a German speaking university in Linz \cite{staudigl}. 

\begin{figure}[!t]
\centering
\includegraphics[width=12.5cm]{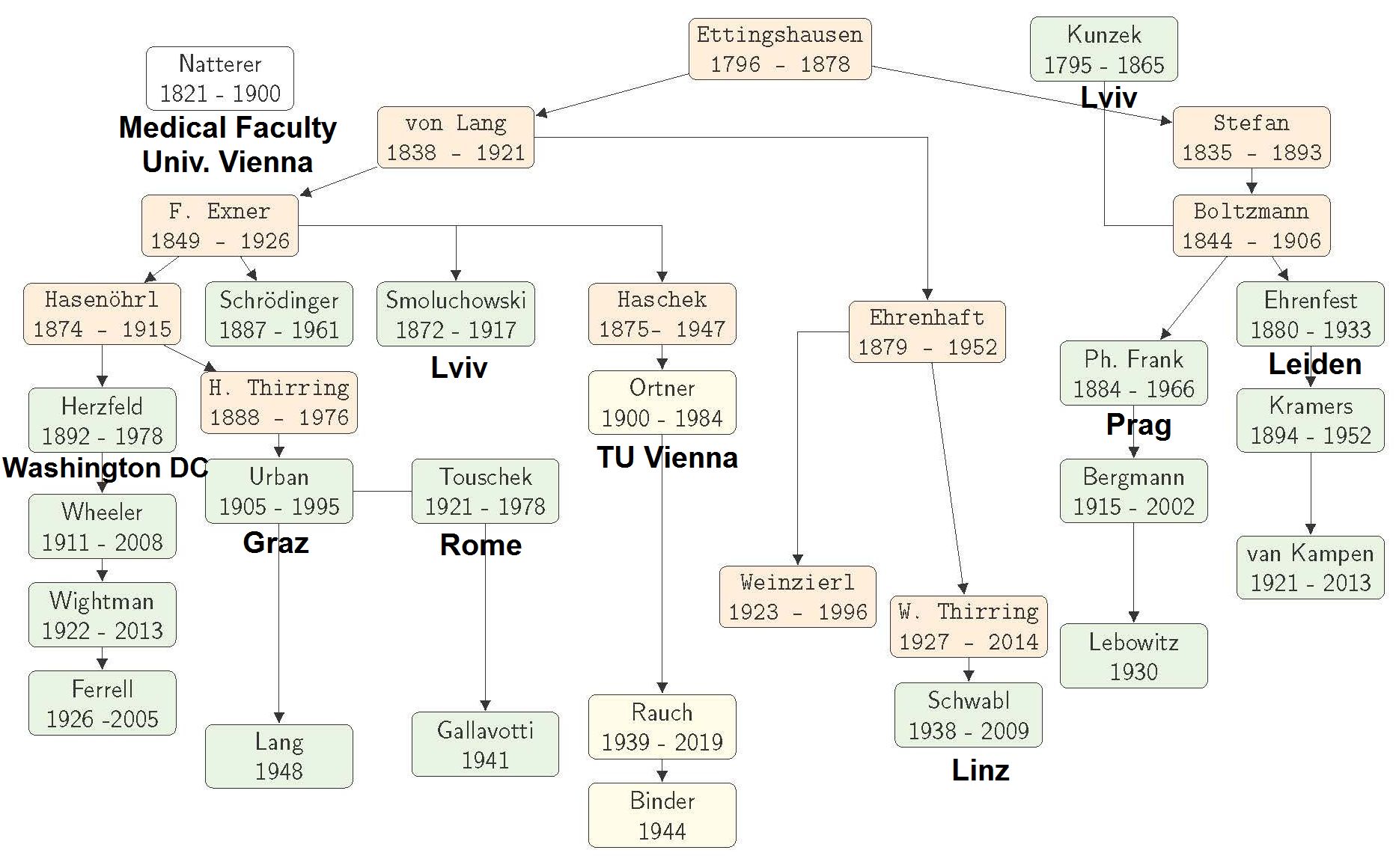}
\caption{\label{fig5a} (Colour online) Part of the genealogical  diagram, according to the relation between PhD supervisor and student 
 indicated by an arrow, of the {\it Vienna School of Statistical Thought}. The line without arrow indicates lecturing,  see text.}
\end{figure}

Statistical physics, a science that uses probabilistic methods in solving physical problems of behaviour of many-particle interacting systems,
as a separate field of research was born in the middle of 19th century. As noted by Mark Kac, a former student of Hugo Steinhaus in Lviv University, 
``About the 
middle of the nineteenth century, attempts were 
begun to unite the disciplines of mechanics and thermodynamics'' \cite{Kac}. 
He named Boltzmann besides Maxwell and Gibbs as roots of these far-reaching achievements in science. Indeed, it is hard to overestimate
the role of Vienna school in creating and developing statistical physics. Elliott W. Montroll noted on the AIP Conference   
{\it Random Walks and Their Applications in the Physical and Biological Sciences} \cite{montroll} in 1982: ``A 
remarkably large number of pioneers in the evolution of the statistical style of thought in modern physical and indeed biological science may trace 
their family 
tree back to the Vienna school\ldots I must confess that not a drop of Wienerblutt {\it (sic!)} flows in my 
veins nor have any of my teachers been leaves on the Vienna Family Tree. However, my life has been very much enriched by my friends 
and colleagues who are leaves identified or not identified on the tree.'' 
Montroll's family tree \cite{ptree}  follows 
some of the roots to the leaves. In figure~\ref{fig5a}, on the one hand, we have corrected\footnote{Corrections are the following: Stefan was not the supervisor of Smoluchowski but Lang and Exner \cite{rovenchak1}.
Formally W. Thirring had Ehrenhaft as supervisor, see Physics/Mathematics Tree and Wikipedia.}, reduced  and, on the other hand, extended the tree with 
respect to the topics of collaboration between the {ITP} and ICMP.   Due to the fate of 
Jewish scientists like Ehrenfest and Herzfeld, the thought of this school was scattered all over the world as far as USA and China. Moreover, 
the period of the 
Nazi-Regime after 1938 further increased the brain drain to other countries.

Many of the scientists named in the tree, contributed to the understanding of critical phenomena in liquids, later to the magnetic phase transition and to the 
understanding of ferromagnetism. Of great importance was the struggle to solve the Ising model suggested 1920 by Wilhelm Lenz\footnote{{Less known is 
that Touschek, a Jewish student from Vienna lived in Nazi-time  undercover in Hamburg in the flat of Lenz and later 1957 taught Di Castro Statistical 
Physics using the textbook of Erwin Schr\"odinger \cite{castro}. Di Castro is one of the first advisory board members of MECO.}} and first attacked by 
Ernst Ising, 
although with a solution only for the one dimensional model. 
However, the success made by Onsager 1944 by the solution of the two dimensional model was based on the results developed by Wannier and Kramers, a 
leave of Montroll's tree. These works considerably advanced the research in critical phenomena and in the related fields, see \cite{Ising17} and references therein.

{Scientists left Vienna and got new positions in other countries/institutions and thus enlarged the branches (indicated at their names).} Let us follow some branches 
in the tree in order to  follow the {\it propagation of the thoughts} and see how they reached Linz and Lviv.

After his thesis 1895, M. Smoluchowski went on a tour through Europe where he worked in Glasgow with Lord Kelvin. After returning to Vienna he 
habilitated\footnote{His talk for this occasion had the title: ``The energy distribution in the spectrum of a black body'' \cite{hoeflechner1,hoeflechner2}.} and 
looked for a permanent position and  got at the end of the year 1899 the chair of Mathematical (Theoretical) Physics at the University in Lviv. 

K.F. Herzfeld\footnote{He was one of the first to comment  E. Ising's paper 1925. In the same year he published 
his book on kinetic theory and statistical mechanics, which became a graduate-level textbook in German-speaking universities, 
see also \cite{johnson}.} after a stay 1919 in Munich left Europe and got a visiting professorship 1926 at the Johns Hopkins University in 
Baltimore, Maryland and 1933 a chair at  The Catholic University of America in Washington, D.C. One of his descents, R.A. Ferrell 
made important contributions related to the further development of statistical physics and critical phenomena and stimulated also 
the connections within this field in Europe. {This also leads to an example for the touch of two different leaves of the tree in later times in 
the coauthor network, when since 1967 a series of ``United Nations papers'' \cite{ferrell} by Ferrell and visitors from Hungary (Nora Menyhard and 
Peter Szepfalusy), Germany (Hartwig Schmidt) and Austria (Franz Schwabl) on the scaling theory in critical dynamics appeared. Some of the authors were 
later involved in organizing the above mentioned MECO conferences. F. Schwabl was appointed  a chair at the University of Linz in 1973 and initiated 
these studies at the ITP.}

P. Ehrenfest left Vienna 1906 and became 1912 after stays in G\"ottingen, St. Peterburg\footnote{{During this stay he connected the Vienna school of 
statistical thought to the Lviv school shown in figure~\ref{lviv} in seminars together with Ioffe. He also published at this time, together with his wife 
Tatyana Afanasyeva, the famous review about statistical physics.}}  the successor of  H.A. Lorentz at the University of 
Leiden {(see \cite{huijen})}. 
There his descendant made important contributions to the understanding of critical phenomena. After he left Vienna,  a meeting with Schr\"odinger 
is reported, where he introduced him to the work of Langevin and Weiss \cite{klein}. This stimulated Schr\"odinger's 
%early  on
 {\it Studien \"uber 
Kinetik der Dielektrika, den Schmelzpunkt, Pyro- und Piezoelektrizit\"at} \cite{schroedinger} where he coined the term {\sc ferroelektrisch} (ferroelectric).
{This type of phase transitions was one of the main topics in the first MECO conference 1974 and the topics, which lead to 
a cooperation with  Prof. Yulian Vysochanskii from the Uzhhorod National University and lead to a common `European' publication \cite{vysbook}. It was at 
the Ukrainian, Polish and East-European Workshop on Ferroelectricity and Phase transitions 1994 
in Uzhhorod--V. Remety  where the authors of this paper met first.}

E. Schr\"odinger considered himself a follower of Ludwig Boltzmann through his teacher Franz Exner and foremost, Fritz (Friedrich) Hasen\"ohrl. 
``\ldots His intention to extend the explanatory range of statistical theory guided his choice of topics'' \cite{joas}. Although Schr\"odinger had the 
opportunity 
to stay at the University of Vienna, he left Austria but returned 1936 to the University of Graz. After the `Anschlu\ss \, 1938' he could not 
stay longer in 
Graz and fled to Dublin, where he stayed until his retirement in 1955 and then returned again to Austria to the University of Vienna.

\subsection{Lviv school of statistical physics}

\begin{figure}[!t]
\centering
\includegraphics[width=12.5cm]{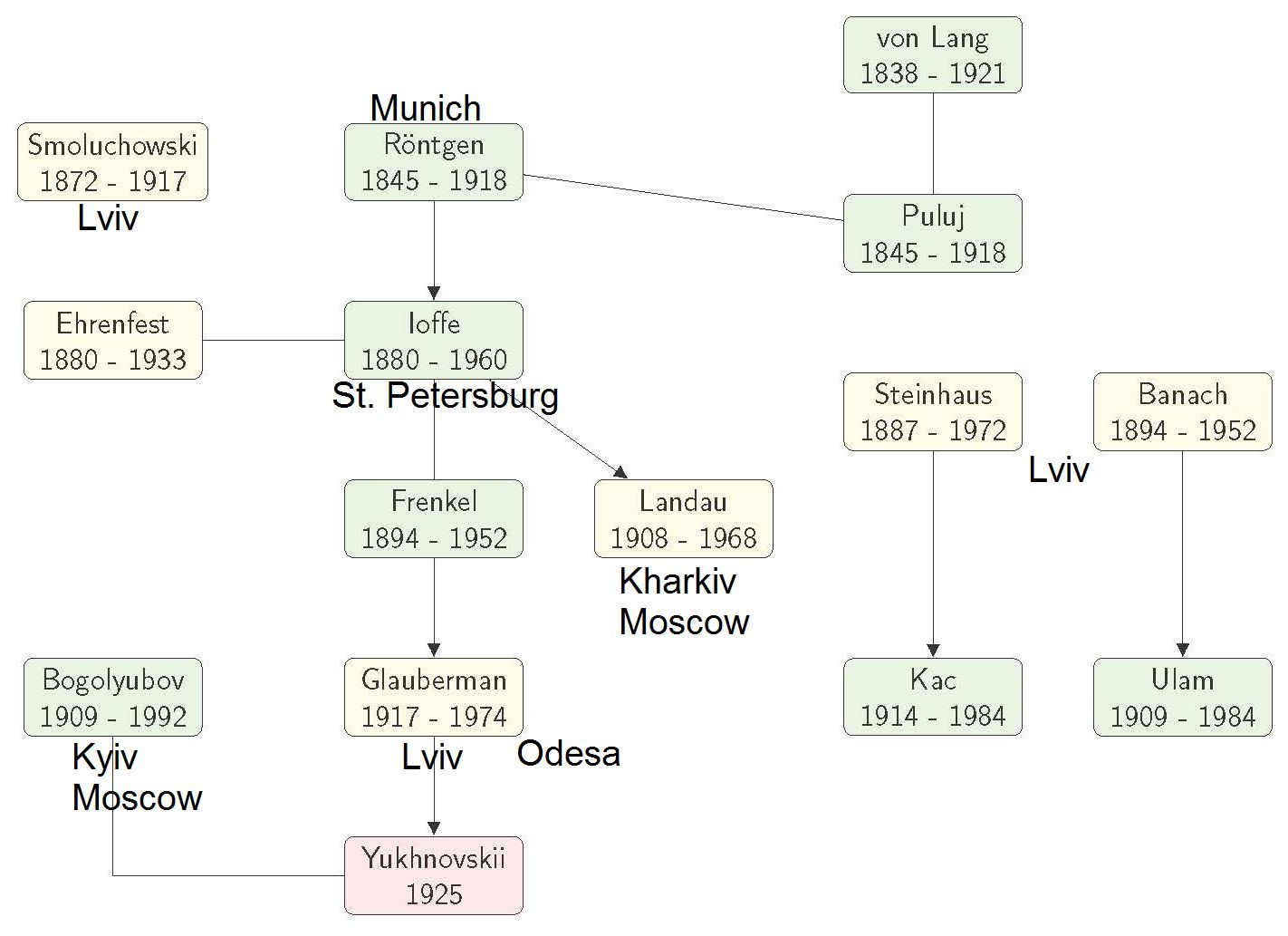}
\caption{\label{lviv} (Colour online) Part of the genealogical  diagram according to the relation between PhD supervisor and student, as indicated by arrows, 
of the {\it Lviv School of Statistical Physics} and the {\it Lviv Mathematical School}.  {The line without arrow indicates other contacts between 
the scientists.}}
  \end{figure}

The roots of theoretical physics in Lviv date back to 1850 \cite{rovenchakKiktyeva}. An important step 
was an appointment of M. Smoluchowski 1899 to the chair of Mathematical (Theoretical) Physics at the University of Lviv~\cite{rovenchak1}. 
In the 1908 paper {\it Molekular-kinetische Theorie der Opaleszenz von Gasen im kritischen Zustande, sowie einiger verwandter Erscheinungen}  
\cite{smoluch1} he linked the critical density fluctuations, {demonstrated  near the  critical point in Natter's tube,} to the fluctuations 
in the refraction index and thus to the scattering in the fluid. Mark Kac described Smoluchowski's scientific output in the following-relying  \cite{kacSM}: 
``\ldots while directed toward the same goal how different the Smoluchowski approach is from Boltzmann's. There is no dynamics, 
no phase space, no Liouville theorem --- in short, none of the usual underpinnings of Statistical Mechanics. Smoluchowski may not 
have been aware of it but he began writing a new chapter of Statistical Physics which in our times goes by the name of 
Stochastic Processes. It is the probabilistic point of view in contradistinction to the statistico-mechanical one that 
is also clearly present in Smoluchowski's first paper on Brownian motion and in a paper on the mean free path which just preceded it.  \ldots
 The underlying idea proved enormously fruitful and it gradually permeated much of statistical physics; permeated it, in fact, so well
 that few of us realize today that much of {\it modern} problematics (notably that related to the so-called master equations) is directly 
 traceable to ideas first promulgated by Smoluchowski in the early years of this century''. 
 Indeed, as it was recognized later it is the appearance of fluctuations and correlations of {\bf all sizes} which 
 make the liquid opaque for all optical wavelength. This `loss of scale' or better invariance of scale at $T_\text{c}$ is fundamental 
 to scaling theory with all its consequences \cite{beysens}.

Stanislaw Ulam characterized the situation of Smoluchovski in Lviv in the following way \cite{ulam}: ``It is interesting to see how it 
was possible for a person of his exceptionally high ability, to get to the forefront of European thought in physics, even though the 
milieu in which he worked as a young professor was relatively isolated and without tradition in science. Nevertheless, it was possible to 
start the pioneering work in a relatively new field (statistical mechanics) and get to the forefront of world science in it, 
{\it once a catalyzing contact with other minds had been made}\footnote{Accentuation by the authors.}''. 
And he immediately compared this with the situation of the {\it Lviv Mathematical School} \cite{Lviv_math1,Lviv_math2} between the  World Wars to which he, as well as
the aforementioned Marc Kac belonged. Activity of the Mathematical School made Lviv a center of several  evolving
fields, functional analysis and set theory being among them. Stanislaw Ulam invented a method that currently has grown up to
a broad class of computational algorithms  widely 
used in physics in general and in statistical physics in particular: in 1940-ies together with John von Neuman he suggested a Monte Carlo method.
{The application to the magnetic transition in the Ising model was presented by K. Binder 1974 at the first MECO conference. 
K. Binder remembers \cite{simu}: ``I was the only person studying Ising systems in Europe at the time and he [David Landau] was the only one doing the same in the United States. So, he read my papers when they appeared and we agreed to meet in a conference, which was a magnetism conference in 1973 in Moscow''.}

In the year 1962  Mark Kac, G.E. Uhlenbeck, and P.C. Hemmer started a series of papers on van der Waals theory to make the theory less qualitative. 
The concept of a  critical {\it region} in contrast to a  critical {\it point} has often been discussed, especially in order to explain
various anomalous critical phenomena which apparently
were in conflict with the van der Waals equation. They also mentioned the effect of gravity on critical phenomena and the attempt of G. Bakker \cite{bakker1,bakker2} 
to deduce a critical region instead of just a critical point. Reasons for the controversial views are the interpretations of the disappearance of 
the meniscus  around the critical temperature observed in Natterer's tube. The shape of the coexistence curve was also under discussion.

Besides Lviv University and Polytechnic, physics has been developing in the Shevchenko Scientific Society --- a prototype of the first Ukrainian National
Academy of Sciences \cite{Lviv_NTSH,Holovatch18}. In this way, the scientists could create what nowadays is called a {\it scientific community}, a community with an own identity 
extending over national and cultural  borders. It was first of all, for theoretical physicist, the time of creating modern quantum mechanics and nuclear physics. 
It was also the time of upcoming 
displacements and strengthening the global identity of the scientists community. A more detailed description of this process is given by M. Desser 
in his book {\it Between Scylla and Charybdis: The `scientific community' of physicists 1919--1939} \cite{desser}.

World War II  changed the academic landscape in Lviv, also in the field of theoretical physics. Ihor Yukhnovskii recalls: ``\ldots Bogolyubov's impact 
on the Lviv school of theoretical physics was induced by his well-known book
{\em Problems of dynamic theory in statistical physics} that was published in 1946. It was a young scientist Abba Glauberman,
a postgraduate from Leningrad (now St. Peterburg), who in 1948 turned our attention upon this book here in Lviv\footnote{{His geneological 
path goes back to Ioffe, who is one of the founders of  theoretical physics in the post-tsarist 
Russia (see figure~\ref{lviv}).  1966 he moved to Odesa \cite{glauberman}. {For the development of theoretical physics of the ``Landau school'' see \cite{hall}. For Frenkel's relation with Ioffe see \cite{tamm}. }}}. ``Having hand-written my candidate\footnote{I. Yukhnovskii PhD thesis entitled 
{{\it Binary distribution function for the systems of interacting charged particles}} has 
been defended under supervisorship
of A. Glauberman \cite{Yukhnovskii_PhD}.} dissertation in a nice big notebook, I brought it to Moscow University to find
the seminar conducted by Bogolyubov\ldots This is how Bogolyubov directly entered my life and the life of theoretical physicists in Lviv''
\cite{Yukhnovskii09}. This was the start of the new period in the development of statistical physics in Lviv, which gave rise
to the ICMP and to the Lviv-Linz collaboration. A detailed description of ICMP organizational and intellectual development
may be found in a recently published book \cite{ICMP_book}.

\section{Instead of conclusions} \label{IV}

This very short and selected history of scientific cooperation between academic institutions in Austria and Ukraine shows, in our 
opinion, several characteristics: (1) Historical roots of cooperation can survive over long periods. (2) Common language 
(Latin in Europe and German in Habsburg Empire in the past and English now) in publications, conferences 
facilitates to easily overcome the cultural differences\footnote{{For the situation in the multilingual academic space in Imperial Austria 
see \cite{surman2}}.}. (3)~Transnational activities by general European programs or created on 
a smaller scale are essential to  intensify and enlarge such cooperations. However, at the beginning of such a process one always 
has a personal decision to leave common paths and try  new adventures in solving physical problems.

So one of us (R.F.) thanks Ihor Mryglod for capturing the opportunity for a fruitful cooperation between ITP and ICMP, 
whereas the other one (Yu.H.) joins his thanks also for interesting discussions and time spent together working and travelling.
We thank Harald Iro and anonymous referees for useful suggestions, Yu.H. acknowledges OeAD scholarship
ICM-2018-11442.

\ukrainianpart

\title{Долаючи кордони у XIX столітті і зараз --- два приклади сплітання наукової мережі}

\author{Р. Фольк\refaddr{label1}, Ю. Головач\refaddr{label2,label3,label4}}
\addresses{
	\addr{label1} Інститут теоретичної фізики, Університет Йоганна Кеплера Лінц, 4040 Лінц, Австрія
	\addr{label2} Інститут фізики конденсованих систем НАН України, 79011 Львів, Україна
	\addr{label3} Співпраця $\mathbb{L}^4$ і Докторський коледж статистичної фізики складних систем,
	Ляйпціґ-Лотаринґія-Львів-Ковентрі, Європа
	\addr{label4} Центр плинних і складних систем, Університет Ковентрі, Ковентрі, CV1 5FB, Велика Британія
}

\makeukrtitle

\begin{abstract}
Наукове дослідження є і було в усі часи транснаціональною (глобальною) діяльністю. У цьому відношенні воно долає кілька кордонів: національний, культурний, ідеологічний. Навіть у часи, коли наукову спільноту розмежовували фізичні кордони, учені тримали розум відкритим для ідей, створених по інший бік мурів і намагались спілкуватися, незважаючи на всі перешкоди. Прикладом такої діяльності в галузі фізики є подорож 1838 року  у західну Європу трьох вчених --- Андреаса Еттінґсгаузена (професора Віденського університету), Августа Кунцека (професора Львівського університету) та о. Маріана Коллера (директора обсерваторії у місті Кремсмюнстер, Верхня Австрія).
155 років пізніше розпочався жвавий науковий обмін між фізиками Австрії та України, зокрема між Інститутом фізики конденсованих систем НАН України у Львові та Інститутом теоретичної фізики університету Йоганна Кеплера в Лінці. Такий обмін став можливим завдяки програмам, що фінансуються національними установами, але він мав свої наукові передумови в уже існуючих наукових мережах, коли Львів був міжнародним центром математики, а у Відні виникла «Школа статистичної думки». Завдяки новій співпраці Україна стала першою державою після розпаду Радянського Союзу, яка приєдналась до ініціативи Середньоєвропейського співробітництва зі статистичної фізики (MECO), заснованої на початку 1970-х років з метою подолання розриву між розділеними залізною завісою вченими східної та західної частин Європи.
У цій статті, обговорюючи наведені вище приклади наукової співпраці, ми ставимо перед собою декілька завдань: зафіксувати менш відомі факти з історії науки в загальному культурологічному контексті, простежити розвиток досліджень, що спричинили появу статистичної фізики та фізики конденсованої речовини, прослідкувати за розвитком багатошарових мережевих структур, що об’єднують науковців уможливлюючи їхні дослідження. Ми із задоволенням подаємо цю статтю у спеціальний випуск журналу, присвячений 60-річчю відомого фізика, нашого доброго колеги та друга Ігоря Мриглода. Своєю працею він зробив значний внесок у зміцнення мереж, про які ми розповідаємо у цій роботі.

\keywords історія науки, історія фізики, статистична фізика

\end{abstract}

\begin{thebibliography}{99}


\bibitem{heisenberg} Heisenberg W., Physics and Beyond, Harper \& Row, New York, 1971. 
%Dresden M., H.A. Kramers Between Tradition and Revolution, Springer-Verlag, New York, 1987.  

\bibitem{kauffman} Kauffman S., \textit{Beyond Physics: an Origin and Evolution of Life}, A talk on December 6, 2017 at
QBio Seminar at the Wisconsin Institute for Discovery, URL~\url{https://vimeo.com/246871926}.

\bibitem{liberwolf} Liberman S., Wolf K.B., In: Handbook of the Psychology of Science, Feist G.J., Gorman M. (Eds.), Springer Publishing Co., New York, 2013, 123--147.

\bibitem{stanley} Zeng A., Shen Z., Zhou J., Wua J., Fan Y., Wanga Y., Stanley H.E., Phys. Rep., 2017, {\bf 714--715}, 1--73, \doi{10.1016/j.physrep.2017.10.001}.

\bibitem{shietal} Shi F., Foster J.G., Evans J.A., Social Networks, 2015, \textbf{43}, 73--85, \doi{10.1016/j.socnet.2015.02.006}.

\bibitem{resnetw} Contandriopoulos D., Larouche C., Duhoux A., Can. J. Program Eval., 2018, \textbf{33}, 69--89, \doi{10.3138/cjpe.42159}.

\bibitem{Berche16} Berche B., Holovatch Yu., Kenna R., Mryglod O., J. Phys. Conf. Ser., 2016, {\bf 681}, 012004,\\ \doi{10.1088/1742-6596/681/1/012004}.


\bibitem{Lviv_math1} Prytula Ya.H.,  Math. Bull. Shevchenko Sci. Soc., 2013, {\bf 10}, 7--16 (in Ukrainian).
\bibitem{Lviv_math2}  Prytula Ya., Banach T., In: Book of Abstracts of the International Conference in Functional Analysis Dedicated to the 125th Anniversary of Stefan Banach (Lviv, 2017), Shevchenko Scientific Society, Lviv, 2017, 13--17,\\ URL~\url{http://at.yorku.ca/c/b/o/m/31.htm}.

\bibitem{Math} Duda R., Lviv Mathematical School, Wroclaw University Publ., Wroclaw, 2007, (in Polish).


\bibitem{montroll} Montroll E.W., AIP Conf. Proc., 1984, {\bf 109}, 1--10, \doi{10.1063/1.34336}.

\bibitem{amand} Kraml P.A., {\it Eine ``physikalische Reise'' durch
Westeuropa im Jahr 1838}, A talk at the 62nd Annual Meeting of the Austrian Physical Society, Graz, 2012.
R.F. thanks P.A.K. for providing the transparencies of the talk.

\bibitem{felloecker} Fell\"ocker P.S., Geschichte der Sternwarte der Benediktiner --- Abrei Chremsminster, J. Feichtinger's Erben, Linz, 1864.  


\bibitem{reslhuber} Reslhuber A., In: Die Feierliche Sitzung der Kaiserlichen Akademie der Wissenschaften am 31 Mai 1867, Wien Hof- und Staatsdruckerei, 1867, 105--143. 

\bibitem{wolf} Wolf R., Rudolf Wolfs Jugendtagebuch 1835--1841, Schriftenreihe der ETH-Bibliothek, Band 30,  1993.   

\bibitem{ditchen1} Ditchen H., Die Politechnika Lwowska in Lemberg. Geschichte einer Technischen Hochschule im multinationalen Umfeld, Logos Verlag Berlin Stuttgarter Beitr\"age zur Wissenschafts- und Technikgeschichte, Band 7,  2015.  

\bibitem{surman} Surman J.J.,  Habsburg Universities 1848--1918. Biography of a Space, Thesis University of Vienna, 2012.  

\bibitem{katgraz} H\"oflechner W. (Hrsg.), Katalog zur Ausstellung {\em Ludwig Boltzmann Anl\"asslich des 100. Todestages} Physik an der Universit\"at Graz, Graz, 2006.   


\bibitem{zamg} Webpage of the Zentralanstalt f\"ur Meteorologie und Geodynamik section `Observatorien' (in German), URL~\url{https://www.zamg.ac.at/cms/de/geophysik/magnetik/observatorien}.

\bibitem{koller} Koller M.,  Ueber den Gang der W\"arme in \"Osterreich ob der Enns, Friedrich Eurich, Linz, 1841.

\bibitem{coen}  Coen D.R.,  Climate in Motion: Science, Empire, and the Problem of Scale, University of Chicago Press, Chicago, 2018.

\bibitem{BHK1} Berche B., Henkel M., Kenna R., Rev. Bras. Ensino Fis., 2009, {\bf  31}, 2602,\\ \doi{10.1590/S1806-11172009000200015}.
\bibitem{BHK2} Berche B., Henkel M., Kenna R., J. Phys. Stud., 2009, {\bf 13}, 3201. 

\bibitem{Holovatch17} Holovatch Yu., Kenna R., Thurner S.,  Eur. J. Phys., 2017, {\bf 38},  023002, \doi{10.1088/1361-6404/aa5a87}.
\bibitem{Ising17} Ising T., Folk R., Kenna R., Berche B., Holovatch Yu.,  J. Phys. Stud., 2017, {\bf 21}, 3002, \doi{10.30970/jps.21.3002}.
\bibitem{almqvist} Almqvist E., History of Industrial Gases, Springer Science \& Business Media, New York, 2003.
\bibitem{brenni} Brenni P., Bull. Sci. Instrum. Soc., 2013, \textbf{117}, 2--8.

\bibitem{price} De Solla Price D.J., Little Science, Big Science, Columbia University Press, New York, 1963.

\bibitem{beaver}  Beaver D.d., Rosen R., Scientometrics, 1979, \textbf{1}, No.~3, 231--245, \doi{10.1007/BF02016308}.
\bibitem{iaria} Iaria A., Schwarz C., Waldinger F., Q. J. Econ., 2018, \textbf{133}, 927--991, \doi{10.1093/qje/qjx046}.
\bibitem{fortunato} Fortunato S., Bergstrom C.T., B\"orner K., Evans J.A., Helbing D., Milojevi\'c S., Petersen A.M., Radicchi F., Sinatra R., Uzzi B., Vespignani A., Waltman L., Wang D., Barab\'asi A.-L., Science, 2018, {\bf 359}, eaao0185, \doi{10.1126/science.aao0185}. 

\bibitem{MECO} Materials about MECO History, URL~\url{https://sites.google.com/site/mecoconferencephysics/home}.

\bibitem{paper} Mryglod I.M., Tokarchuk M.V., Folk R., Physica A, 1995, \textbf{220}, 325, \doi{10.1016/0378-4371(95)00232-V}.

\bibitem{Iro_book} 
 Iro H., Klassische Mechanik, Universit\"atsverlag Rudolf Trauner, Linz, 1996, (in German) [Iro H., Klasychna Mekhanika,  
Ivan Franko National University of Lviv, Lviv, 1997, (in Ukrainian);
 Iro H., A Modern Approach to Classical Mechanics, World Scientific, Singapore, 2015].

\bibitem{staudigl} Staudigl-Ciechowicz K., Zwischen Wien und Czernowitz --- \"Osterreichische Universit\"aten um 1918, In: Beitr\"age zur Rechtsgeschichte \"Osterreichs, Band 2, 2014, 223--240.


\bibitem{Kac} Kac M.,  Statistical Independence in Probability Analysis and Number Theory, Mathematical Association of America, Washington, DC, 1959.  


%%%Development%%%%%%%%%%%%%%%%%%%%

\bibitem{ptree} The Physics Tree, URL~\url{https://academictree.org/physics/}.

\bibitem{rovenchak1} Rovenchak A., Trokhymchuk A., Math. Model. Comput., 2018, {\bf 5}, 99--107, \doi{10.23939/mmc2018.02.099}.

 \bibitem{castro} Di Castro C., Bonolis L., Eur. Phys. J. H, 2014, {\bf 39}, 3--36, \doi{10.1140/epjh/e2013-40043-5}.

\bibitem{hoeflechner1} H\"oflechner W., Die Physik an der Universit\"at Wien 1848--1938, Unver\"offentliches Typoskript, 1992. 

\bibitem{hoeflechner2} H\"oflechner W., Materialien zur Entwicklung der Physik und ihrer Randf\"acher Astronomie und Meteorologie an den \"Osterreichischen Universit\"aten 1752--1938,
Unver\"offentlichtes Typoskript, Graz, 2002,\\ URL~\url{http://www-gewi.uni-graz.at/wissg/geschichte_der_physik/index.html}.  

\bibitem{johnson} Johnson K.E., J. Stat. Phys., 1990, {\bf 59}, 1547--1572, \doi{10.1007/BF01334763}.

\bibitem{ferrell} Sucher J., Scalapino D., Prange R., Phys. Today, 2006, {\bf 59}, 71, \doi{10.1063/PT.4.2323}.

\bibitem{huijen} Huijen P., Kox A.J., Phys. Perspect., 2007, {\bf 9}, 186--211, \doi{10.1007/s00016-006-0287-1}.

\bibitem{klein} Klein M.J., Paul Ehrenfest, North-Holland Publishing Company, Amsterdam, London, 1970.

\bibitem{schroedinger} Schr\"odinger E., Studien \"uber Kinetik der Dielektrika, der Schmelzpunkt, Pyro- und Piezoelektrizit\"at, Sitzungsberichte der kaiserlichen Akademieder
Wissenschaften in Wien. Mathematisch-Naturwissenschaftliche Klasse~(IIa), 1912, \textbf{121}, 1937--1972.  

\bibitem{vysbook}  Vysochanskii Yu.M., Janssen T., Currat R., Folk R., Banys J., Grigas J., Samulionis V., Phase Transitions in Ferroelectric Phosphorous Chalcogenide Crystals, Vilnius University Publishing House, Vilnius, 2006.
 
\bibitem{joas} Joas Ch., Katzir S., Stud. Hist. Philos. Sci., Part B, 2011, {\bf 42}, 43--53, \doi{10.1016/j.shpsb.2010.12.004}.


%%%%%%%%%%%%%%%%%%Lviv school %%%%%%
\bibitem{rovenchakKiktyeva} Rovenchak A., Kiktyeva O., Studia Historiae Scientiarum, 2016, {\bf 15}, 48--73,\\ \doi{10.4467/23921749SHS.16.004.6147}.

\bibitem{smoluch1} Smoluchowski M.v., Ann. Phys., 1908, {\bf 330}, 205--226, \doi{10.1002/andp.19083300203}.

\bibitem{kacSM} Kac M., In: Marian Smoluchowski, His Life and Scientific Work, Polish Scientific Publishers PWN, Warszawa, 1999.

\bibitem{beysens} Beysens D.A., In:  Mechanics of the 21st Century,  Gutkowski W., Kowalewski~T.A.~(Eds.), Springer, Dordrecht, 2005. 

\bibitem{ulam} Ulam S., Am. J. Phys., 1957, {\bf 25}, 475--481, \doi{10.1119/1.1934510}.

\bibitem{simu} Mac Kernan D., Interview of Kurt Binder, In: SIMU Challenges in Molecular Simulations: Bridging the Length- and Timescales Gap, Vol. 3, 2001, 7--31. 


\bibitem{bakker1} Bakker G.,  Ann. Phys., 1904, {\bf 320}, 543--553, \doi{10.1002/andp.19043201307}.
\bibitem{bakker2} Bakker G.,  Z. Phys. Chem., 1904, {\bf 49}, 609--617, \doi{10.1515/zpch-1904-0146}.
%%%%%%%%%%%%%%%%%%%%%%%%%%%%

\bibitem{Lviv_NTSH} Kravtsiv B., Kubijovy\u{c} V., Shevchenko Scientific Society, In: Internet Encyclopedia of Ukraine;
and the on-going multivolume series Encyclopedia of the Schevchenko Scientific Society, Lviv, 2013-- (in Ukrainian).  

\bibitem{Holovatch18} Holovatch Yu., Honchar J., Krasnytska M.,  J. Phys. Stud., 2018, {\bf 22}, 4003 (in Ukrainian), \\ \doi{10.30970/jps.22.4003}. 

\newpage

\bibitem{desser} Desser M., Zwischen Skylla und Charybdis: Die `Scientific Community' der Physiker 1919--1939, B\"ohlau  Verlag, Wien, K\"oln, Weimar, 1991. 

\bibitem{glauberman}  Talyanskii  I., J. Phys. Stud., 1998, {\bf 2}, 150--153 (in Ukrainian). 

\bibitem{hall} Hall K., In: Intelligentsia Science: The Russian Century, 1860--1960, University of Chicago Press, Chicago, 2008, 230--259.  

\bibitem{tamm} Tamm I.E.,  Sov. Phys. Usp., 1962, \textbf{5}, 173--194, \doi{10.1070/PU1962v005n02ABEH003406}. 

\bibitem{Yukhnovskii_PhD} Binary Distribution Function for the Systems of Interacting Charged Particles, (Based on PhD Thesis by I.R. Yukhnovskii), Eurosvit, Lviv, 2010, (in Ukrainian). 

\bibitem{Yukhnovskii09} Yukhnovskii I., Condens. Matter Phys., 2009, \textbf{12}, No. 4, 535--537, \doi{10.5488/CMP.12.4.535}.

\bibitem{ICMP_book} Mryglod~I.~(Ed.), Institute for Condensed Matter Physics of the National Academy of Sciences of Ukraine: the Half-Century Trip, 
 ICMP of the National Academy of Sciences of Ukraine, Lviv, 2019, (in Ukrainian).

\bibitem{surman2} Surman J.J., Universities in Imperial Austria, 1848--1918: A Social History of a Multilingual Space, Purdue University Press Books, West Lafayette, Indiana, 2019.

\end{thebibliography}
\end{document}